\pdfoutput=1
\documentclass[conference]{IEEEtran}
\IEEEoverridecommandlockouts
\usepackage{cite}
\usepackage{amsmath,amssymb}
\usepackage{booktabs,array,multirow}
\usepackage{graphicx}
\usepackage{xcolor}
\definecolor{flowBlue}{HTML}{246A99}
\definecolor{nativeTeal}{HTML}{157A6E}
\definecolor{fabricAmber}{HTML}{B66C16}
\definecolor{routeSlate}{HTML}{57667C}
\usepackage{tikz}
\usetikzlibrary{positioning,arrows.meta,calc,patterns}
\usepackage{pgfplots}
\pgfplotsset{compat=1.18}
\usepgfplotslibrary{groupplots}
\usepackage{xspace}
\usepackage[hidelinks]{hyperref}
\usepackage{tabularx}
\hypersetup{pdfauthor={Mohamed Bouaziz, Mahmoud A. Elshenawy, Krishnan B. Iyer, Abdurauf Abdurakhmanov, Ammar M. Seliaman, Suhaib A. Fahmy},pdftitle={Native FPGA Hard-Block Implementation from High-Level Synthesis}}
\newcommand{\ours}{Ours\xspace}
\newcommand{\X}{$\times$}
\newcommand{\factor}[2]{\textnormal{(#1$\times\!#2$)}}
\newcommand{\ns}{\,ns\xspace}
\begin{document}
\title{Native FPGA Hard-Block Implementation from High-Level Synthesis}
\author{\IEEEauthorblockN{Mohamed Bouaziz, Mahmoud A. Elshenawy, Krishnan B. Iyer, Abdurauf Abdurakhmanov,\\ Ammar M. Seliaman, and Suhaib A. Fahmy}
\IEEEauthorblockA{King Abdullah University of Science and Technology (KAUST), Thuwal, Saudi Arabia}}
\IEEEaftertitletext{\vspace{-1.0\baselineskip}}
\maketitle
\begin{abstract}
High-level synthesis (HLS) maps computations to FPGA hard blocks, yet often leaves their arithmetic modes, internal registers, and cascade links idle and recreates them in the fabric, costing area and timing. We present a C++ template library that reaches native hard-block implementations without modifying the HLS compiler. Templates fix source patterns, and an automatic flow picks scheduling settings and verified RTL patches. On AMD Versal, our matrix multiplication uses 2.4--4.5$\times$ fewer lookup tables (LUTs) than AutoSA systolic arrays at equal or higher estimated throughput. Our INT8 matrix--vector units reach the three multiply--accumulates per DSP of hand-written FINN RTL, versus at most 1.29 for HLS libraries. In hls4ml, Allo, and Stream-HLS, the templates cut LUTs by up to 2.7$\times$ and DSPs by up to 2$\times$.
\end{abstract}
\begin{IEEEkeywords}
High-level synthesis, FPGA synthesis, technology mapping.
\end{IEEEkeywords}
\section{Introduction}
\label{sec:intro}

High-level synthesis (HLS) lets developers build field-programmable gate array (FPGA) accelerators from C++ instead of register-transfer-level (RTL) code. Pragmas control pipelining, parallelism, and storage, while the compiler binds operations to resources~\cite{cong2011hls,ug1399}. In particular, it maps arithmetic to digital signal processing (DSP) blocks and arrays to block RAM (BRAM) and UltraRAM (URAM). These hard blocks offer more than a multiplier or a memory array. They also provide pre-adders, wide accumulators, single-instruction multiple-data (SIMD) and floating-point modes, optional internal pipeline registers, and dedicated cascade links that pass operands and partial sums to neighboring blocks without general routing~\cite{boutros2021arch}. Recent device families extend this set further, with INT8 dot products and single-precision floating-point (FP32) and complex arithmetic in the DSPs, and configurable cascades in the memories~\cite{versalacap2019,am004,am007}.

\begin{figure*}[t]
\centering
\begin{tikzpicture}[x=1mm,y=1mm,font=\footnotesize,
 box/.style={draw=flowBlue,fill=flowBlue!7,rounded corners=1pt,
   minimum width=13mm,minimum height=6mm,align=center,inner sep=1.5pt},
 ours/.style={box,draw=nativeTeal,fill=nativeTeal!12,thick},
 arr/.style={-{Stealth[length=1.3mm]},draw=routeSlate,thick},
 ctl/.style={-{Stealth[length=1.3mm]},draw=fabricAmber,dashed,thick},
 cas/.style={-{Stealth[length=1.3mm]},draw=nativeTeal,very thick},
 note/.style={font=\scriptsize,align=center},
 dsp/.style={draw=flowBlue,fill=flowBlue!12,minimum width=11mm,minimum height=8mm,align=center,inner sep=1pt},
 fab/.style={draw=fabricAmber,dashed,fill=fabricAmber!14,align=center,inner sep=1pt},
 reg/.style={draw=nativeTeal,fill=nativeTeal!25,minimum width=2.6mm,minimum height=2.6mm,inner sep=0pt,font=\tiny},
 mem/.style={draw=flowBlue,fill=flowBlue!12,minimum width=15mm,minimum height=11mm,align=center,inner sep=1pt},
 ttl/.style={font=\footnotesize\bfseries,anchor=west}]
 \path[use as bounding box] (-1,-7) rectangle (180,50);
 \draw[routeSlate!40] (-1,24.5) -- (181,24.5);
 \node[ttl] at (-1,47.5) {(a) Conventional HLS};
 \node[ttl] at (-1,21.5) {(b) \ours};
 \node[font=\footnotesize\itshape,text=routeSlate] at (40,47.5) {flow};
 \node[font=\footnotesize\itshape,text=routeSlate] at (137,47.5) {resulting mapping};
 \node[box,minimum width=24mm] (s0) at (14,37) {C++ + pragmas};
 \node[box] (h0) at (41,37) {HLS};
 \node[box] (r0) at (60,37) {RTL};
 \node[box] (i0) at (79,37) {Vivado};
 \foreach \a/\b in {s0/h0,h0/r0,r0/i0} \draw[arr] (\a) -- (\b);
 \node[ours,minimum width=24mm] (s1) at (14,11) {C++ with\\[-1pt]our templates};
 \node[box] (h1) at (41,11) {HLS};
 \node[box] (r1) at (60,11) {RTL};
 \node[box] (i1) at (79,11) {Vivado};
 \foreach \a/\b in {s1/h1,h1/r1,r1/i1} \draw[arr] (\a) -- (\b);
 \node[note,text=nativeTeal] at (14,0) {source patterns\\[-1pt](Sec.~\ref{sec:mapping})};
 \node[note,text=fabricAmber] (u) at (41,-1.5) {clock uncertainty,\\[-1pt]op.\ latencies\\[-1pt](Sec.~\ref{sec:auto})};
 \node[note,text=fabricAmber] (p) at (61,-1.5) {verified\\[-1pt]RTL patches\\[-1pt](Sec.~\ref{sec:auto})};
 \draw[ctl] (u) -- (h1);
 \draw[ctl] (p.north) -- (r1.south);
 \draw[arr] (i0) -- (94,37);
 \node[mem] (m0) at (103,37) {};
 \fill[flowBlue!35] (m0.south west) rectangle ($(m0.south west)+(15,3)$);
 \node[font=\scriptsize,align=center] at (103,38.5) {BRAM\\[-1pt]partly used};
 \foreach \i/\x in {1/119,2/133,3/147} {
   \node[dsp] (a\i) at (\x,41) {DSP\\[-1pt]$\times$};
   \node[fab,minimum width=6mm,minimum height=3mm,font=\tiny] (f\i) at (\x,32.5) {FF};
   \draw[arr] (a\i) -- (f\i);
 }
 \draw[arr] (m0.east) -- (a1.west);
 \node[fab,circle,minimum size=4.5mm] (t1) at (126,27.5) {$+$};
 \node[fab,circle,minimum size=4.5mm] (t2) at (140,27.5) {$+$};
 \draw[arr,fabricAmber] (f1) -- (t1); \draw[arr,fabricAmber] (f2) -- (t1);
 \draw[arr,fabricAmber] (f2) -- (t2); \draw[arr,fabricAmber] (f3) -- (t2);
 \draw[arr,fabricAmber] (t2.east) -- (157,27.5);
 \node[note,text=fabricAmber,anchor=west,align=left] at (156,36) {fabric:\\[-1pt]adder trees,\\[-1pt]registers,\\[-1pt]LUT logic};
 \draw[arr] (i1) -- (94,11);
 \node[mem,draw=nativeTeal,fill=nativeTeal!12] (m1) at (103,11) {};
 \foreach \k in {0,...,3} \draw[nativeTeal!60] ($(m1.south west)+(0,2.2+2.2*\k)$) -- ++(15,0);
 \node[font=\scriptsize,align=center,fill=nativeTeal!12,inner sep=0.5pt] at (103,11) {BRAM\\[-1pt]9$\times$INT8/word};
 \foreach \i/\x in {1/119,2/133,3/147} {
   \node[dsp,draw=nativeTeal,fill=nativeTeal!8,minimum height=11mm] (b\i) at (\x,11) {};
   \node[font=\scriptsize] at (\x,13.6) {$\times\,+$};
   \node[reg] at ($(b\i.south)+(-2.4,2.6)$) {A};
   \node[reg] at ($(b\i.south)+(2.4,2.6)$) {P};
 }
 \draw[arr] (m1.east) -- (b1.west);
 \draw[cas] (b1.east) -- (b2.west);
 \draw[cas] (b2.east) -- (b3.west);
 \draw[cas] (b3.east) -- (157,11);
 \node[note,text=nativeTeal,anchor=west,align=left] at (156,13) {hard blocks:\\[-1pt]modes, internal\\[-1pt]registers, cascades,\\[-1pt]packed memories};
 \node[note,text=nativeTeal] at (133,1.5) {cascade links};
\end{tikzpicture}
\caption{Overview. (a) Conventional HLS rebuilds hard-block capabilities in the fabric. (b) Our templates, scheduling settings, and verified RTL patches keep them inside the hard blocks.}
\label{fig:overview}
\end{figure*}
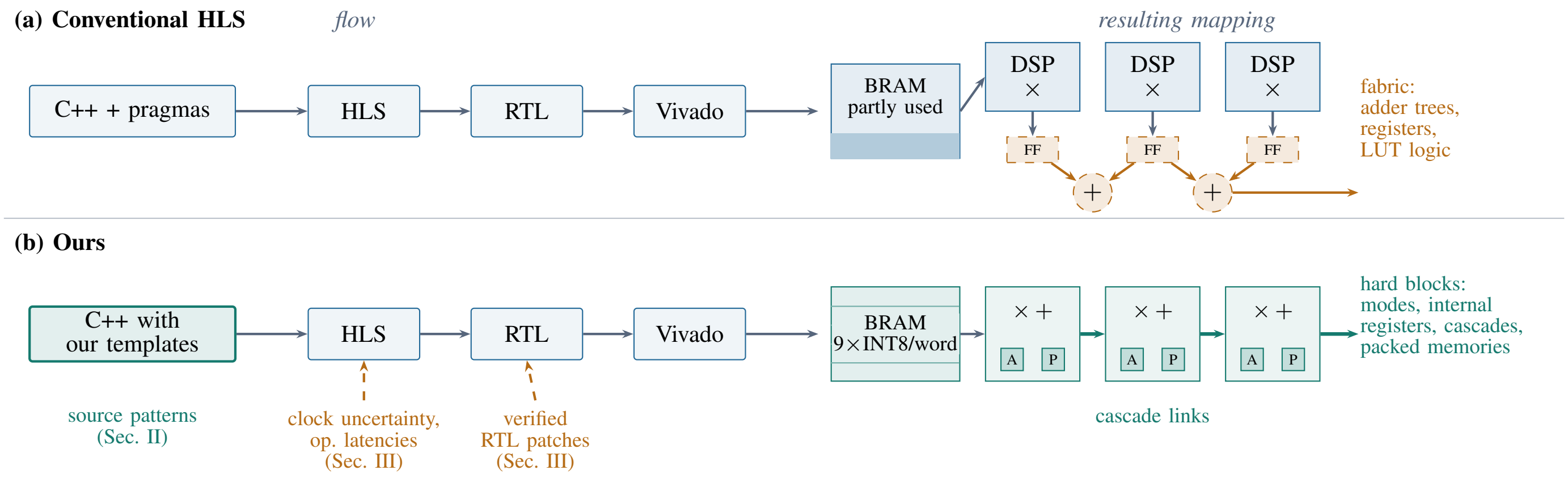

\begin{figure}[t]
\centering
\begin{tikzpicture}[
  x=1mm,y=1mm,font=\scriptsize,
  dsp/.style={
    draw=flowBlue,fill=flowBlue!12,
    minimum width=18mm,minimum height=6mm,
    align=center,inner sep=1pt
  },
  fab/.style={
    draw=fabricAmber,dashed,fill=fabricAmber!12,
    minimum width=11mm,minimum height=5mm,
    align=center,inner sep=1pt
  },
  arr/.style={
    -{Stealth[length=1.3mm]},draw=routeSlate
  },
  native/.style={
    arr,draw=nativeTeal,thick
  },
  bus/.style={
    draw=flowBlue,thick
  }
]
\path[use as bounding box] (-0.5,-2) rectangle (88,43);
\begin{scope}
  \node[font=\scriptsize\bfseries,anchor=west]
    at (0,42) {(a) Plain HLS};

  \draw[bus] (2,0) -- (2,34);
  \node[anchor=south west] at (2,34)
    {$x$ on \text{B}};

  \foreach \i/\y in {0/28,1/19,63/3} {
    \node[dsp] (a\i) at (15,\y)
      {$h_{\i}x+z_{\i}$\\[-1pt]
       \text{A1}: $h_{\i}$};

    \node[fab] (f\i) at (34,\y)
      {delay\\[-1pt]SRL/FF};

    \draw[arr] (2,\y) -- (a\i.west);

    \draw[arr] (f\i.west) -- (a\i.east);
  }

  \node at (15,11) {$\vdots$};
  \node at (34,11) {$\vdots$};

  \draw[arr,dashed]
    (a1.north) -- (a0.south);

  \draw[arr] (a0.north)
    -- ++(0,4) node[above] {$y$};

  \draw[arr] (15,-2.5)
    -- (a63.south);
  \node[font=\tiny,anchor=north] at (15,-3)
    {48-bit bias on \text{C}};

  \node[
    align=left,anchor=north west,
    text width=41mm
  ] at (0,-5.5)
    {Fabric-assisted acc.
     and external delays};
\end{scope}

\begin{scope}[xshift=45mm]
  \node[font=\scriptsize\bfseries,anchor=west]
    at (0,42) {(b) \ours};

  \draw[
    fill=nativeTeal!6,
    draw=nativeTeal!55
  ] (6,-1) rectangle (26,32);

  \draw[bus] (2,0) -- (2,34);
  \node[anchor=south west] at (2,34)
    {$x$ on \text{B}};

  \foreach \i/\y in {0/28,1/19,31/3} {
    \node[dsp] (b\i) at (16,\y)
      {$h_{\i}x+\text{PCIN}$\\[-1pt]
       \text{A1}: $h_{\i}$, \text{P}: $z$};

    \draw[arr] (2,\y) -- (b\i.west);
  }

  \draw[native]
    (b1.north) -- (b0.south);

  \draw[native]
    (b31.north) -- (16,8);

  \node[
    fill=nativeTeal!6,
    inner sep=1pt
  ] at (16,11) {$\vdots$};

  \draw[native]
    (16,14) -- (b1.south);

  \node[
    font=\tiny,align=left,
    anchor=west,text=nativeTeal
  ] at (26.5,11)
    {\text{PCOUT}\\$\to$\text{PCIN}};

  \draw[native] (b0.north)
    -- ++(0,4) node[above] {$y$};

  \draw[arr] (16,-2.5)
    -- (b31.south);
  \node[font=\tiny,anchor=north] at (16,-3)
    {48-bit bias on \text{C}};

  \node[
    align=left,anchor=north west,
    text width=41mm
  ] at (0,-5.5)
    {Native acc. and registers;
     $2\!\times\!32$ taps};
\end{scope}

\end{tikzpicture}
\caption{A 64-tap FIR filter mapped by (a) plain HLS and (b) our library.}
\label{fig:fir}
\end{figure}
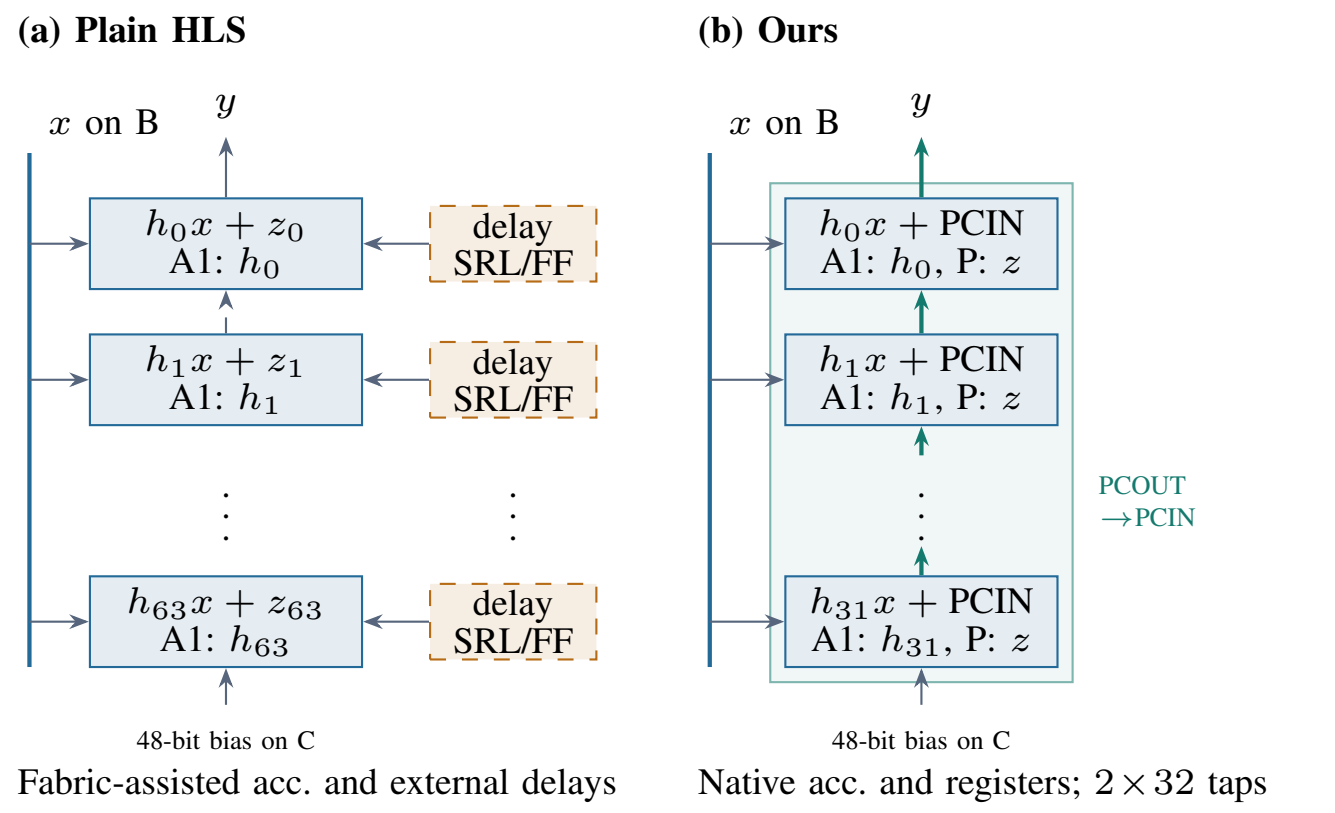

However, inferring a hard block does not mean its capabilities are fully utilized. Rather, HLS often rebuilds some of these in the fabric, as Fig.~\ref{fig:overview}a illustrates. We call a mapping \emph{native} when arithmetic, state, and communication stay inside the hard blocks, through their modes, internal registers, and cascade links. Fig.~\ref{fig:fir} shows a 64-tap finite impulse response (FIR) filter, $y[n]=\sum_{k=0}^{63} h_k\,x[n-k]$. Both versions use 64 DSPs, one per coefficient $h_k$. In plain HLS (Fig.~\ref{fig:fir}a), the compiler sums the products with an adder tree and cascades only the pairs at its bottom level. As a result, it forms 32 of the 62 possible cascade links, and the delayed partial sums $z_k$ move to fabric shift registers (SRLs) and flip-flops (FFs). The filter then needs 1008 lookup tables (LUTs), 1437 SRLs, and 5086 FFs around its DSPs. In contrast, the native mapping (Fig.~\ref{fig:fir}b) chains the DSPs in two columns of 32. Each DSP multiplies, adds the partial sum arriving from its neighbor (PCIN), keeps the delay in its own result register (P), and passes the sum on (PCOUT). Only 70 LUTs remain, to load the coefficients. Such extra fabric costs area and, because it sits on the paths between blocks, complicates timing closure as designs grow~\cite{tapa2023}.

Existing approaches do not close this gap from HLS. Hand-written RTL libraries, such as FINN's matrix--vector units~\cite{finn2017,finnrepo}, reach native mappings but depart from general purpose HLS. HLS libraries and frameworks, such as finn-hlslib~\cite{finnhlslib}, hls4ml~\cite{hls4ml2018}, AutoSA~\cite{autosa2021}, Allo~\cite{allo2024}, and Stream-HLS~\cite{streamhls}, decide loops, dataflow, and schedules, but leave binding to the compiler and therefore inherit its mapping. Finally, research on better DSP mapping targets RTL generators or earlier devices, and arithmetic only~\cite{ronak2016tcad,dsppacking2022}. We therefore ask the following research question. \textbf{Can HLS code reach the native mapping of FPGA hard blocks without changing the HLS compiler?}

We answer it with a C++ template library for FPGA hard blocks, summarized in Fig.~\ref{fig:overview}b. The library addresses three issues. The first issue is that HLS selects a hard-block configuration only for specific source patterns. Therefore, each template fixes the C++ expression and pragmas that produce one configuration, found by inspecting placed designs. The second issue is that HLS scheduling decides where pipeline registers go. In particular, the clock uncertainty and operator latencies decide whether a register stays inside a DSP or moves to the fabric. Thus, an automatic flow sets them from per-operator timing budgets. The third issue is that HLS cannot express some configurations, such as accumulation in the DSP result register. For these, the flow applies small RTL patches after HLS and keeps only those that verifiably improve the design. Our contributions are as follows:
\begin{itemize}
\item \textbf{A native mapping library.} Templates for memories, integer, INT8, SIMD, FP32, and complex arithmetic, cascades, lanes, and matrix tiles achieve native configurations of the AMD Versal hard blocks from HLS (Section~\ref{sec:mapping}).
\item \textbf{An automatic mapping flow.} It selects clock uncertainty, operator latencies, and RTL patches without modifying the compiler (Section~\ref{sec:auto}).
\item \textbf{An evaluation at three levels.} We compare isolated operations with plain HLS, complete kernels with FINN RTL, finn-hlslib, and AutoSA, and integrations into hls4ml, Allo, and Stream-HLS (Section~\ref{sec:eval}).
\end{itemize}
On a Versal VC1902, our matrix multiplication uses 2.4--4.5\X{} fewer LUTs than the leanest AutoSA systolic array at equal or higher estimated throughput. Moreover, its INT8 matrix--vector units reach three multiply--accumulates (MACs) per DSP, as FINN RTL does, whereas HLS libraries reach at most 1.29.

\section{The Template Library}
\label{sec:mapping}

Each template of our library is a C++ function, with explicit width, timing, and control parameters, whose body HLS maps to one hard-block configuration. This section describes which configurations the library covers, how we found the source pattern for each, and the RTL patches that complete the configurations HLS cannot express.

\begin{table}[t]
\centering
\caption{Template families and OOC costs
(300\,MHz unless marked).}
\label{tab:families}

\small
\setlength{\tabcolsep}{0pt}
\renewcommand{\arraystretch}{1.08}

\begin{tabularx}{\columnwidth}{
@{}
>{\raggedright\arraybackslash}X
@{\hspace{7pt}}
>{\raggedleft\arraybackslash}p{.35\columnwidth}
@{}
}
\toprule
\textbf{Template / configuration} &
\textbf{Typical cost} \\
\midrule

\multicolumn{2}{@{}l@{}}{
\textbf{Memories}
} \\
\multicolumn{2}{@{}l@{}}{
\footnotesize\emph{Cascade height; write mode; URAM ROM}
} \\[1pt]

BRAM/URAM, ROMs, packed/banked arrays and buffers
& 1 block/array, 0 LUT/FF \\

URAM (up to 32K words)
& 0 LUT; deeper arrays cascade \\

\addlinespace[5pt]
\multicolumn{2}{@{}l@{}}{
\textbf{Integer DSP}
\quad\footnotesize\emph{Clearable sum}
} \\[1pt]

27\X24 multiply, multiply--add, pre-add, 58-bit add
& 1 DSP, 0 LUT \\

Accumulator
& 1 DSP, 1 LUT, 4 FF \\

Accumulator with CLR
& 1 DSP, 0 LUT, 3 FF \\

48\X48 multiply
& 5 DSP, 0 LUT \\

64/96-bit accumulator
& 3 DSP, 0 LUT \\

\addlinespace[5pt]
\multicolumn{2}{@{}l@{}}{
\textbf{Cascades}
} \\[1pt]

64-tap FIR
& 64 DSP, 70 LUT, 0 FF \\

Result-cascaded and memory-fed dot products
& 1 B18/tap, 0 LUT/SRL \\

\addlinespace[5pt]
\multicolumn{2}{@{}l@{}}{
\textbf{INT8 and SIMD}
\quad\footnotesize\emph{SIMD; logic}
} \\[1pt]

Three-product INT8 dot
& 1 DSP, 0 LUT \\

At 800\,MHz
& 1 DSP, 1 LUT \\

2\X24- or 4\X12-bit add
& 1 DSP, 0 LUT \\

\addlinespace[5pt]
\multicolumn{2}{@{}p{\columnwidth}@{}}{
\textbf{Floating point}\quad
\footnotesize\emph{Uncertainty;
operand gating}
} \\[1pt]

FP32 add, subtract, multiply and multiply--add
& 1 \text{DSP}, at most 2 LUT \\

FP32 complex multiply
& 4 \text{DSP} \\

\multicolumn{2}{@{}l@{}}{
Dot products and sums at II=1
} \\

\addlinespace[5pt]
\multicolumn{2}{@{}l@{}}{
\textbf{Complex Numbers}
\quad\footnotesize\emph{Adder merge}
} \\[1pt]

18\X18 complex multiply
& 2 \text{DSP}, 0 LUT/FF \\

Multiply--add and two-term dot product
& 0 LUT \\

\addlinespace[5pt]
\multicolumn{2}{@{}l@{}}{
\textbf{Lanes and pairs}
} \\[1pt]

$L$ lanes (up to 64 integer/FP32 or 32 complex)
& $L$ blocks, 0 LUT/lane \\

Memory--DSP pairs
& 0 LUT/FF \\

One clock region
& 24 B36 + 96 DSP \\

\addlinespace[5pt]
\multicolumn{2}{@{}l@{}}{
\textbf{Tiles}
\quad\footnotesize\emph{Clearable sum}
} \\[1pt]

INT16 GEMV (cascaded, up to 128 DSP)
& 0 LUT/DSP \\

4\X4 matrix multiply with local sums
& 16 DSP, 1 LUT, 4 FF \\

With CLR
& 16 DSP, 0 LUT, 3 FF \\

INT8 tiles
& 12.7--15.8 LUT/DSP \\

\bottomrule
\end{tabularx}

\par\vspace{3pt}
\begin{minipage}{\columnwidth}
\footnotesize\raggedright
Italic entries identify RTL patches or required settings.
CLR: explicit clear input; B18/B36: 18/36-Kbit BRAM.
One complex operation uses two DSPs.
\end{minipage}
\end{table}

\subsection{Target Configurations}

The templates follow the capabilities of modern FPGA hard blocks~\cite{am004,am007}, grouped into the families of Table~\ref{tab:families}. For DSPs, these capabilities are the arithmetic modes, accumulator widths, internal registers, and cascade links. For memories, they are the widths and depths, operating modes, and cascade links. Lanes replicate a template up to 64 times, and tiles compose templates into matrix units, such as a 4\X4 matrix multiply whose 16 DSPs keep their sums locally. Most templates leave at most a few LUTs and FFs around their blocks. Some accumulators also take an explicit clear input (CLR), which restarts the sum inside the DSP. To use a template, a developer replaces an operator or an array in existing C++ code with a template call, and the rest of the design, including its pragmas, stays unchanged.

\subsection{Finding the Source Patterns}

For each template, we explored C++ expressions and pragmas until HLS produced the intended configuration. Each candidate was synthesized by HLS and then synthesized and placed out of context (OOC). We then inspected the primitive types and counts, the enabled modes and internal registers, the cascade links, the remaining fabric logic, and the timing slack, and kept the candidates that matched the intended mapping. For example, the FIR template of Fig.~\ref{fig:fir}b expresses the filter as a chain in which each tap adds its product to the registered sum of the previous tap, rather than as one sum over all taps. Most templates reach their mapping with pragmas alone.

\subsection{RTL Patches}

Some remaining templates need a configuration that the compiler does not expose. For these, we patch the RTL after HLS. \emph{Attribute patches} change a primitive setting without changing the surrounding logic. For example, raising the URAM cascade height above its default builds deeper memories through native cascades. \emph{Logic patches} change the generated datapath or control. For example, one patch replaces an external accumulator register with the DSP's internal result register. The italic entries in Table~\ref{tab:families} mark the families that need a patch or a scheduling setting. Each patch is written for the RTL structure that HLS generates for its template. If the surrounding code leads HLS to generate a different structure, the patch may not apply or may break the design. Therefore, the flow in the next section checks every patch before applying it.

\section{Automatic Mapping Flow}
\label{sec:auto}

A template fixes what HLS should build, but the HLS scheduler still decides where pipeline registers go. If the scheduler estimates that a DSP's output path does not fit in one cycle, it adds a fabric register after the block instead of using the block's internal one. Moreover, some configurations cannot be reached from C++ at all. Therefore, the flow in Fig.~\ref{fig:overview}b acts at two points without modifying the compiler. Before HLS, it selects the clock uncertainty and operator latencies. After HLS, it selects and verifies RTL patches.

\subsection{Clock Uncertainty and Operator Latency}

HLS schedules each cycle against a time budget $B=T-u$, where $T$ is the target clock period and $u$ is the clock uncertainty, a margin reserved for implementation delays. By default, $u_d=0.27T$. However, this margin also changes the generated design. In a hard block, operands move through dedicated internal paths rather than via general-purpose fabric routing, which needs almost no margin and therefore less uncertainty.  When $B$ is shorter than the internal path of a DSP mode, HLS adds an external result register, for example after an FP32 adder or multiply--add. We therefore measured, for each operator class $c$, the budget $B_c$ at which the native mapping appears. It is 2.83\ns for FP32 additions and multiply--adds, and 3.33\ns for FP32 multiplications and conversions. The flow then sets
\begin{equation}
u_c=\begin{cases} T-B_c & \text{if } T-u_d<B_c\leq T,\\ u_d & \text{otherwise.}\end{cases}
\label{eq:uncertainty}
\end{equation}
In other words, \eqref{eq:uncertainty} lowers the uncertainty just enough when the default budget is too short. At slower clocks, the default budget already suffices, and at faster clocks ($T<B_c$), no non-negative uncertainty reaches $B_c$. For example, at 300\,MHz, $T=3.33$\ns and the default budget is 2.43\ns, below the 2.83\ns that an FP32 adder needs. Thus, the flow sets $u=0.50$\ns, or 15\% of $T$. Complex-number operators behave in the opposite way. Their DSP mode enables all internal pipeline registers only when $B\leq1.43$\ns, so the flow raises the uncertainty to $u_c=\max(u_d,\,T-B_c)$.

A lower uncertainty can also let HLS shorten an operator's pipeline and drop internal DSP registers. Hence, for FP32 operators at reduced uncertainty, the flow also tries fixed latencies taken from a table of measured operator configurations. Operators absent from the table keep the default uncertainty.

The selection proceeds as follows. The flow first synthesizes the design with $u_d$ and identifies its operators. Without FP32 or complex operators, it keeps $u_d$. Otherwise, it builds the candidates above and rejects those in which an operator has a lower latency than in the default synthesis, unless a measurement confirms that it still meets $T$. Among the candidates that meet timing, it selects the one with the lowest resource cost and the most native mappings. In case of equality, it prefers $u_d$, and otherwise the uncertainty closest to $u_d$. If no candidate meets timing, the flow reports the one with the highest worst negative slack (WNS) and its unmet margin. Finally, because the uncertainty applies to a whole kernel, operators with conflicting requirements cannot both be served. In that case, the flow recommends splitting them into separate kernels.

\subsection{Patch Selection and Verification}

After scheduling, the flow takes each patch in its registry (Section~\ref{sec:mapping}) and dry-runs it on a copy of the RTL. It then checks the conditions that the patch relies on in the source and in the HLS reports, for example that an accumulator loop never stalls. A patched design must build and keep the intended primitive types. The flow accepts the patch if it lowers the resource cost and either meets timing or loses at most 0.1\ns of slack, which admits a slightly negative WNS. An attribute patch that leaves the cost unchanged still qualifies if it shortens the achievable clock period by more than 0.05\ns.

\section{Experimental Results}
\label{sec:eval}

We evaluate whether our library reaches native mappings in isolation, how close it comes to hand-written RTL and generated accelerators, and whether existing frameworks benefit from it.

\subsection{Setup}

We implemented our library for AMD Versal, whose enhanced DSPs support INT8, FP32, and complex arithmetic, and whose BRAMs and URAMs enable configurable cascades~\cite{versalacap2019,am004,am007}. Experiments target the VC1902 device on the VCK5000 board~\cite{vck5000} using Vitis HLS and Vivado 2022.1 at 300\,MHz. Each table and figure states whether results come from OOC placement or from post-route kernels. Throughput values are estimates at the target clock. \emph{Plain HLS} denotes the same computation written in C++ with standard pragmas and without our library. We compare at three levels. First, isolated operations against plain HLS. Second, complete kernels against FINN RTL~\cite{finn2017} and finn-hlslib~\cite{finnhlslib}, and against systolic arrays generated by AutoSA~\cite{autosa2021}. Third, integrations into hls4ml~\cite{hls4ml2018}, Allo~\cite{allo2024}, and Stream-HLS~\cite{streamhls}, using their own workloads, some from PolyBench~\cite{polybench}.

Some baselines were written for earlier devices, so we adapted them to the Versal. For AutoSA, we updated its storage pragmas and selected its configurations from HLS measurements instead of its UltraScale+ resource predictor. For hls4ml, we restored its disabled DSP pipeline registers to the HLS default. For Stream-HLS, we set the cost of an FP32 addition or multiplication to 1 DSP and the budget to the device's 1968 DSPs.

\subsection{Native Operations}
\label{sec:ooc}

\begin{figure}[t]
\centering
\begin{tikzpicture}[
  plain/.style={draw=fabricAmber,fill=fabricAmber!35,postaction={pattern=north east lines,pattern color=fabricAmber}},
  ours/.style={draw=nativeTeal,fill=nativeTeal!70}]
\begin{groupplot}[
  group style={group size=2 by 2, horizontal sep=7mm, vertical sep=7.5mm, x descriptions at=edge bottom},
  ybar, scale only axis, width=35.5mm, height=17mm, /pgf/bar width=1.55mm,
  ymin=0, enlarge x limits=0.07,
  symbolic x coords={IM8,W48,S8,MACc,M4c,D32,F8,FD8}, xtick=data,
  xticklabel style={font=\tiny,rotate=90,anchor=east,xshift=-0.3mm,yshift=0mm},
  yticklabel style={font=\tiny,/pgf/number format/1000 sep={}}, scaled y ticks=false,
  /pgf/number format/1000 sep={},
  title style={font=\scriptsize,yshift=-1.8mm},
  nodes near coords, every node near coord/.append style={font=\fontsize{4.2}{4.2}\selectfont,rotate=90,anchor=west,xshift=-0.3mm,inner sep=0.3pt},
  axis line style={routeSlate}, tick style={routeSlate},
]
\nextgroupplot[title=DSP, ymax=68, ytick={0,25,50},
  legend to name=nativeleg, legend columns=2, legend style={font=\scriptsize,draw=none,column sep=2mm}]
\addplot[plain] coordinates {(IM8,15) (W48,4) (S8,0) (MACc,1) (M4c,16) (D32,48) (F8,8) (FD8,9)};
\addplot[ours] coordinates {(IM8,8) (W48,5) (S8,8) (MACc,1) (M4c,16) (D32,32) (F8,8) (FD8,9)};
\legend{plain HLS, \ours}
\nextgroupplot[title=LUT, ymax=2050, ytick={0,1000}]
\addplot[plain] coordinates {(IM8,233) (W48,186) (S8,1401) (MACc,49) (M4c,769) (D32,1239) (F8,466) (FD8,772)};
\addplot[ours] coordinates {(IM8,0) (W48,0) (S8,0) (MACc,0) (M4c,0) (D32,0) (F8,9) (FD8,170)};
\nextgroupplot[title=FF, ymax=1550, ytick={0,500,1000}]
\addplot[plain] coordinates {(IM8,43) (W48,130) (S8,346) (MACc,55) (M4c,775) (D32,1075) (F8,728) (FD8,379)};
\addplot[ours] coordinates {(IM8,30) (W48,201) (S8,29) (MACc,3) (M4c,3) (D32,378) (F8,38) (FD8,378)};
\nextgroupplot[title=WNS (ns), ymax=2.4, ytick={0,1,2},
  nodes near coords={\pgfmathprintnumber[fixed,fixed zerofill,precision=2]{\pgfplotspointmeta}}]
\addplot[plain] coordinates {(IM8,1.513) (W48,0.631) (S8,0.412) (MACc,1.380) (M4c,1.260) (D32,0.506) (F8,1.257) (FD8,0.501)};
\addplot[ours] coordinates {(IM8,1.268) (W48,1.600) (S8,0.878) (MACc,1.416) (M4c,1.469) (D32,1.644) (F8,0.972) (FD8,0.742)};
\end{groupplot}
\node[anchor=south] at ($(group c1r1.north)!0.5!(group c2r1.north)+(0,2.6mm)$) {\ref{nativeleg}};
\end{tikzpicture}
\vspace{-1mm}
\par{\footnotesize\raggedright IM8: 8-lane INT16 multiply; W48: 48\X48 multiply; S8: 8-lane 4\X12-bit SIMD add; MACc: MAC with CLR; M4c: INT16 4\X4 matrix multiply with CLR; D32: INT16 dot, 32 MAC; F8: 9-lane FP32 multiply--add; FD8: FP32 dot, 8 MAC.\par}
\vspace{-1mm}
\caption{Primitive mappings after OOC placement at 300\,MHz (all II $=1$).}
\label{fig:ooc}
\end{figure}

Native mapping removes nearly all fabric around the hard blocks, whereas its effect on DSP count and timing depends on the operation. Fig.~\ref{fig:ooc} shows that all six integer operations drop to zero LUTs, and the two FP32 operations to 9 and 170 LUTs. The DSP count falls when a native mode replaces several blocks. For example, IM8 maps each INT16 multiplication to one DSP (15 to 8), and D32 keeps its whole sum on one result cascade (48 to 32 DSPs). Conversely, it rises when a DSP absorbs fabric logic. W48 uses one more DSP for its native partial-product cascade, and S8 moves SIMD additions from 1401 LUTs into eight DSPs. Internal registers explain the FF savings. MACc and M4c accumulate in the DSP result register (55 to 3 and 775 to 3 FFs), and in F8 the uncertainty from Section~\ref{sec:auto} removes the external result registers (728 to 38 FFs).

However, equal DSP counts and II do not imply an equally native mapping. In FD8, plain HLS reaches the same DSPs and II by interleaving partial sums, which costs 772 LUTs instead of 170. Native mapping also tends to improve timing. The WNS rises in six of the eight operations, for example from 0.51 to 1.64\ns in D32, as fewer paths leave the hard blocks. It drops only in IM8 and F8, yet stays above 0.97\ns.

\subsection{Native INT8 Packing Against RTL}

\begin{table}[t]
\centering
\caption{FINN layers after OOC placement at 300\,MHz.}
\label{tab:oocfinn}
\small\setlength{\tabcolsep}{2.4pt}
\begin{tabular*}{\columnwidth}{@{\extracolsep{\fill}}lrrrrrrr@{}}
\toprule
Version & DSP (INT8) & B36/U & M/D & LUT & FF & C & II \\
\midrule
\multicolumn{8}{@{}l}{\emph{INT8 matrix--vector 4\X9: 36 MAC/cycle}} \\
plain HLS & 28 (4) & 4/0 & 1.29 & 224 & 452 & 12 & \textbf{1} \\
finn-hlslib & 20 (8) & 4/0 & 0.60 & 2980 & 7033 & 4 & 3 \\
FINN RTL & \textbf{12} (\textbf{12}) & 4/0 & \textbf{3.00} & 190 & 387 & 8 & \textbf{1} \\
\ours & \textbf{12} (\textbf{12}) & 4/0 & \textbf{3.00} & \textbf{152} & \textbf{144} & 0 & \textbf{1} \\
\midrule
\multicolumn{8}{@{}l}{\emph{INT8 matrix--vector 16\X18: 288 MAC/cycle}} \\
plain HLS & 224 (32) & 32/0 & 1.29 & 2096 & 3792 & 96 & \textbf{1} \\
finn-hlslib & 224 (48) & 32/0 & 1.29 & 6581 & 16311 & 96 & \textbf{1} \\
FINN RTL & \textbf{96} (\textbf{96}) & 32/0 & \textbf{3.00} & \textbf{574} & \textbf{1226} & 80 & \textbf{1} \\
\ours & \textbf{96} (\textbf{96}) & 32/0 & \textbf{3.00} & 1520 & 2244 & 0 & \textbf{1} \\
\midrule
\multicolumn{8}{@{}l}{\emph{INT8 3\X3 convolution: 288 MAC/cycle}} \\
plain HLS & 224 (48) & 0/0 & 1.29 & 5025 & 10389 & 96 & \textbf{1} \\
finn-hlslib & 184 (58) & 0/0 & 0.78 & 9212 & 19854 & 66 & 2 \\
FINN RTL & \textbf{96} (\textbf{96}) & 0/0 & \textbf{3.00} & \textbf{3287} & 8695 & 80 & \textbf{1} \\
\ours & \textbf{96} (\textbf{96}) & 32/4 & \textbf{3.00} & 4833 & \textbf{7013} & 0 & \textbf{1} \\
\midrule
\multicolumn{8}{@{}l}{\emph{INT16 matrix--vector 8\X8: 64 MAC/cycle}} \\
plain HLS & \textbf{72} (\textbf{0}) & 16/0 & \textbf{0.89} & 280 & \textbf{270} & 40 & \textbf{1} \\
finn-hlslib & \textbf{72} (\textbf{0}) & 15/0 & \textbf{0.89} & 5313 & 13254 & 40 & \textbf{1} \\
\ours & \textbf{64} (\textbf{0}) & 16/0 & \textbf{1.00} & \textbf{8} & \textbf{290} & 56 & \textbf{1} \\
\bottomrule
\end{tabular*}
\par\vspace{2pt}
\begin{minipage}{\columnwidth}\footnotesize\raggedright DSP: total, and in parentheses the number in INT8 mode (three 8-bit multiply-accumulates per DSP); B/U: BRAM/URAM; M/D: MAC per DSP per cycle; C: cascade links. LUTs used as memory are excluded; in the convolution, plain HLS, finn-hlslib, and FINN RTL keep their buffers in 315, 2328, and 2299 such LUTs.
\end{minipage}
\end{table}

Our library reaches the DSP packing of hand-written RTL from HLS, with only a modest fabric overhead. FINN~\cite{finn2017,finnr2018} provides quantized neural-network operators both in RTL~\cite{finnrepo} and in HLS~\cite{finnhlslib}. Its RTL is written by experts and already uses INT8 packing, cascades, and accumulation in internal registers. It is therefore the gold standard for what a native mapping can reach. In Table~\ref{tab:oocfinn}, each processing element (PE) receives the same 512 weight words per memory as our corresponding tile.

The INT8 mode of the DSP performs three 8-bit MACs per cycle. Plain HLS and finn-hlslib reach at most 1.29 MACs per DSP per cycle, and finn-hlslib also misses an II of one in two of the three INT8 configurations. In contrast, FINN RTL and our library reach 3.00, with the same DSP count and an II of one in all three INT8 configurations. Even without packing, in the INT16 case, ours needs 64 DSPs instead of 72 and 8 LUTs instead of 280, with 56 cascade links instead of 40.

Matching FINN-RTL from HLS costs little. FINN RTL accumulates in DSP result registers through result cascades, whereas our INT8 templates accumulate partial sums in fabric. Even so, ours uses less fabric than FINN RTL at the small size (152 vs.\ 190 LUTs) and about a thousand more LUTs and FFs at the 16\X18 size. For convolution, FINN RTL uses fewer LUTs, while ours uses fewer FFs and keeps its buffers in 32 BRAMs and four URAMs instead of LUTs. Reaching the arithmetic efficiency of expert RTL at this fabric cost, while keeping the productivity of HLS, is a favorable trade-off.

\subsection{Matrix Multiplication Against Systolic Arrays}

\begin{table}[t]
\centering
\caption{INT16 GEMM: post-route kernel results at 300\,MHz.}
\label{tab:autosa}
\small\setlength{\tabcolsep}{3pt}
\begin{tabular*}{\columnwidth}{@{\extracolsep{\fill}}lrrr@{}}
\toprule
\multicolumn{4}{@{}l}{\emph{Peak: 256 MAC/cycle $\rightarrow$ 153.6 GOPS}} \\
Metric & \ours & A-L & A-C \\
PE shape & 16T & 2\X4/S32 & 8\X8/S4 \\
\midrule
LUT & \textbf{9466} & \factor{2.4}{\uparrow} 22759 & \factor{7.4}{\uparrow} 70022 \\
FF & \textbf{13021} & \factor{3.8}{\uparrow} 50046 & \factor{8.5}{\uparrow} 110370 \\
B36/B18/U & 50/134/8 & 138/22/0 & 406/66/0 \\
P/A/B & 0/0/0 & \textbf{128/0/0} & \textbf{128/0/0} \\
GOPS & (100\%) \textbf{153.6} & (95.4\%) 146.6 & (99.3\%) 152.5 \\
\midrule
\multicolumn{4}{@{}l}{\emph{Peak: 1536 MAC/cycle $\rightarrow$ 921.6 GOPS}} \\
Metric & \ours & A-L & A-C \\
PE shape & 96T & 4\X12/S32 & 12\X16/S8 \\
\midrule
LUT & \textbf{22385} & \factor{4.5}{\uparrow} 101536 & Implementation \\
FF & \textbf{24792} & \factor{8.6}{\uparrow} 214024 & failed: \\
B36/B18/U & 50/774/24 & 438/82/0 & it requires \\
P/A/B & 0/0/0 & 768/0/0 & 998 BRAMs \\
GOPS & (99.8\%) \textbf{919.7} & (97.7\%) 900.1 & (967 available) \\
\bottomrule
\end{tabular*}
\par\vspace{2pt}
\begin{minipage}{\columnwidth}\footnotesize\raggedright A-L/A-C: leanest/classic AutoSA; parentheses: AutoSA cost relative to ours, or share of peak for the estimated GOPS. T: 4\X4 tile; S: SIMD width; P/A/B: result/operand cascade links.
\end{minipage}
\end{table}

Broadcasting operands to DSPs that accumulate in place costs far less logic than a systolic array with the same MAC rate. AutoSA~\cite{autosa2021} generates systolic arrays from loop nests. Both designs compute $C=AB^{T}$ on 1536\X1536 matrices with INT16 inputs, 32-bit results, and one 512-bit port per matrix. INT16 is the widest power-of-two width that fits one DSP multiplier, so every MAC maps to one DSP in both designs. Thus, the comparison isolates how operands reach the DSPs. We compare at equal peak rates of 256 and 1536 MACs per cycle, with the same clock, ports, and tools. Our design is a grid of pragma-only 4\X4 output-stationary tiles. Each memory read is broadcast to the four DSPs of a tile row or column, and each DSP keeps its sum in its result register. For AutoSA, we build a classic layout (A-C) with many small PEs, and the leanest layout (A-L), which has the fewest LUTs among 11 tested legal configurations and uses fewer PEs with 32-lane dot products.

Table~\ref{tab:autosa} shows that our design uses 2.4--4.5\X{} fewer LUTs and 3.8--8.6\X{} fewer FFs than A-L, and reaches 99.8--100\% of peak throughput instead of the 95.4--97.7\% of A-L. At the larger size, A-C does not fit, because it needs 998 BRAMs while only 967 are available. The difference lies in communication. AutoSA passes operands between PEs through FIFOs rather than through operand cascades, and it sums partial dot products over result cascades and fabric adder trees. Moreover, its systolic tiles spend cycles filling and draining, which lowers throughput. Both designs must also keep data on chip. AutoSA holds it in FIFOs and ping-pong buffers spread across its PEs, whereas our design places it in BRAMs and URAMs next to the DSPs that consume it. Keeping data in memory blocks rather than in the fabric is itself a native use of the hard blocks. In 36-Kbit equivalents, ours even needs fewer BRAMs (117 vs.\ 149 and 437 vs.\ 479), complemented by 8 and 24 URAMs.

\subsection{Mapping Neural-Network Layers}

\begin{figure}[t]
\centering
\begin{tikzpicture}
\begin{axis}[
  xbar, scale only axis, width=0.74\columnwidth, height=33mm, bar width=1.9mm,
  xmin=0, xmax=4.6, xtick={0,0.5,1,1.5,2,2.5,3},
  xlabel={\ours relative to hls4ml (hls4ml $=1$)}, xlabel style={font=\scriptsize,yshift=1.5mm},
  ytick=data, y dir=reverse, enlarge y limits=0.09,
  symbolic y coords={DSP,LUT,FF,BRAM,Cycles,MAC/DSP/cycle},
  yticklabel style={font=\scriptsize}, xticklabel style={font=\scriptsize},
  legend style={font=\scriptsize,at={(0.98,0.98)},anchor=north east,draw=none,fill=white,fill opacity=0.85,text opacity=1},
  legend cell align=left, reverse legend,
  nodes near coords, point meta=explicit symbolic,
  every node near coord/.append style={font=\tiny,anchor=west,xshift=0.3mm,fill=white,inner sep=0.4pt},
  axis line style={routeSlate}, tick style={routeSlate},
  extra x ticks={1}, extra x tick labels={}, extra x tick style={grid=major,grid style={dashed,routeSlate}},
]
\addplot[draw=fabricAmber,fill=fabricAmber!35,postaction={pattern=north east lines,pattern color=fabricAmber}] coordinates {
  (1.000,DSP)[32$\to$32] (0.825,LUT)[1837$\to$1515] (0.753,FF)[2780$\to$2092]
  (1.778,BRAM)[4.5$\to$8] (1.000,Cycles)[64$\to$64] (1.000,MAC/DSP/cycle)[1.00$\to$1.00]};
\addplot[draw=nativeTeal,fill=nativeTeal!70] coordinates {
  (0.500,DSP)[32$\to$16] (0.369,LUT)[8131$\to$3003] (0.577,FF)[8078$\to$4658]
  (0.167,BRAM)[33$\to$5.5] (0.665,Cycles)[15155$\to$10085] (3.013,MAC/DSP/cycle)[0.80$\to$2.41]};
\legend{16-bit dense ($R=64$),8-bit convolution ($R=72$)}
\end{axis}
\end{tikzpicture}\par
\vspace{-3.5mm}
\caption{Neural-network layers from hls4ml (post-route, 300\,MHz).}
\label{fig:hls4ml}
\end{figure}
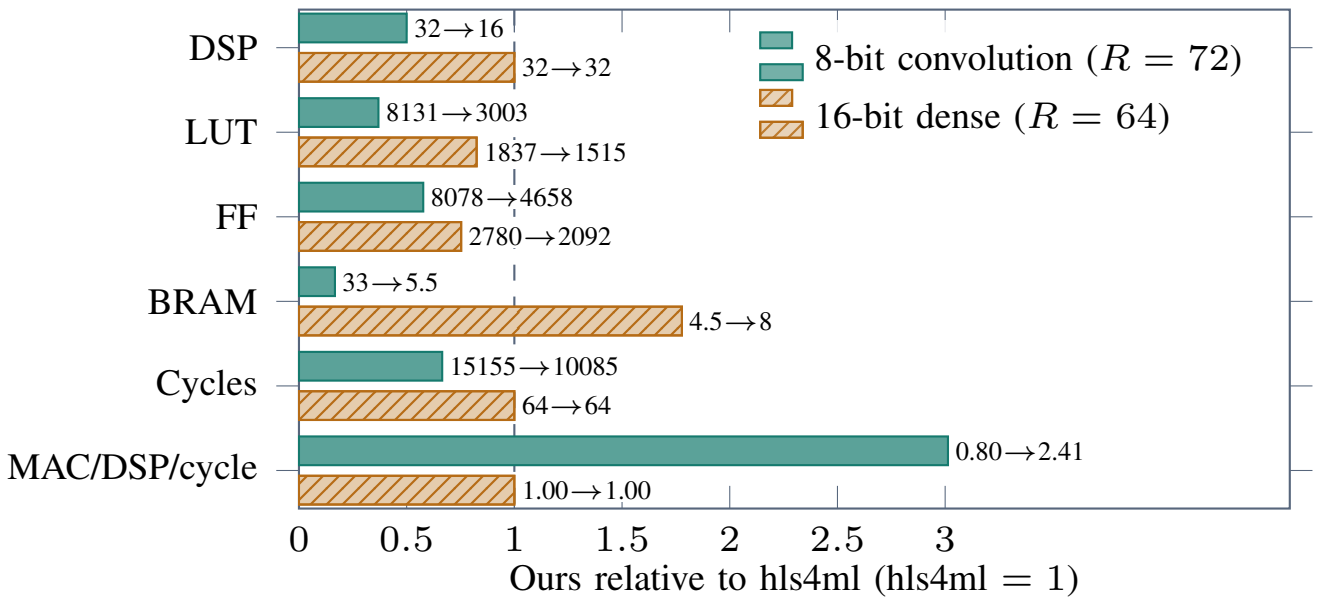

Inside an existing framework, our library helps most where the framework leaves a native capability unused. We evaluate it on two neural-network layers generated by hls4ml~\cite{hls4ml2018}. Fig.~\ref{fig:hls4ml} uses a 64-input, 32-output dense layer from a jet-tagging network~\cite{hls4ml2018} and a 3\X3 convolution with 16 input and output channels on a 15\X15 input~\cite{aarrestad2021}. Both use the `resource' strategy of hls4ml, in which the reuse factor $R$ sets how many times each multiplier is reused. We replace only the dot-product kernel, so the convolution keeps the streaming interface and line buffer of hls4ml.

In the 8-bit convolution, INT8 packing raises the MACs per DSP per cycle from 0.80 to 2.41, which halves the DSPs and cuts cycles by 1.5\X. More importantly, native mapping extends beyond arithmetic to the memory blocks. Packing nine 8-bit weights into each 72-bit word matches the native BRAM width. As a result, it fills the capacity that the shallow, wide weight ROMs of hls4ml leave unused, and needs 6\X{} fewer BRAMs. In contrast, the 16-bit dense layer of hls4ml already accumulates in the DSP. Here, our library keeps the same DSPs and cycles and trades memory depth for logic. It splits the weights into shallower BRAM banks (8 instead of 4.5), and in return reduces LUTs by 1.2\X{} and FFs by 1.3\X.

\subsection{Integration Into Accelerator Generators}

Templates drop into generated designs without changing their schedules, provided that the schedule meets the templates' assumptions. We first integrate our library into Allo, keeping its loops, pragmas, buffers, and interfaces and replacing only operations. In the packed-INT8 systolic GEMM of Allo, written for an older DSP, each PE computes absolute values and restores the signs in logic. Our pre-adder template instead computes the signed products inside the DSP. As a result, it reduces LUTs by 14\% (19,652 to 16,834) and FFs by 20\% (16,917 to 13,611) across 128 PEs, with the same 128 DSPs and schedule. Similarly, in the FP32 GEMM, the multiply--add template with the uncertainty of Section~\ref{sec:auto} keeps the DSP schedule but removes the surrounding registers, which reduces FFs by 38\% (305 to 190) for 17 more LUTs.

The integration fails when the schedule breaks a template's assumption. In the INT8 4\X4 systolic GEMM of Allo, a handshake starts and stops each tile's loop, so the loop can stall. However, the accumulator template assumes a free-running pipeline. Thus, the accumulation falls back to the fabric. With the same DSPs, LUTs grow by 27\% (2,135 to 2,711) and latency by 19\% (5,333 to 6,321 cycles).

\begin{figure}[t]
\centering
\begin{tikzpicture}
\begin{groupplot}[
  group style={group size=3 by 1, horizontal sep=3mm, y descriptions at=edge left},
  ybar, scale only axis, width=22mm, height=22mm, /pgf/bar width=2.1mm,
  ymin=0, ymax=1.62, ytick={0,0.5,1},
  symbolic x coords={GEMM,2mm,FFN}, xtick=data, enlarge x limits=0.25,
  xticklabel style={font=\scriptsize}, yticklabel style={font=\scriptsize},
  title style={font=\scriptsize,yshift=-1.5mm},
  nodes near coords, every node near coord/.append style={font=\tiny,rotate=90,anchor=west,xshift=-0.2mm,fill=white,inner sep=0.4pt},
  nodes near coords={\pgfmathprintnumber[fixed,fixed zerofill,precision=2]{\pgfplotspointmeta}},
  axis line style={routeSlate}, tick style={routeSlate},
  extra y ticks={1}, extra y tick labels={}, extra y tick style={grid=major,grid style={dashed,routeSlate}},
]
\nextgroupplot[title=DSP, ylabel={rel.\ to Stream-HLS}, ylabel style={font=\scriptsize,yshift=-1mm},
  legend to name=streamleg, legend columns=2, legend style={font=\scriptsize,draw=none,column sep=2mm}]
\addplot[draw=nativeTeal,fill=nativeTeal!70] coordinates {(GEMM,0.868) (2mm,0.856) (FFN,0.885)};
\addplot[draw=fabricAmber,fill=fabricAmber!35,postaction={pattern=north east lines,pattern color=fabricAmber}] coordinates {(GEMM,1.210) (2mm,1.284) (FFN,0.959)};
\legend{+\ours (same schedule), +\ours (re-solved)}
\nextgroupplot[title=LUT]
\addplot[draw=nativeTeal,fill=nativeTeal!70] coordinates {(GEMM,0.699) (2mm,0.698) (FFN,0.924)};
\addplot[draw=fabricAmber,fill=fabricAmber!35,postaction={pattern=north east lines,pattern color=fabricAmber}] coordinates {(GEMM,1.061) (2mm,1.070) (FFN,0.822)};
\nextgroupplot[title=Cycles]
\addplot[draw=nativeTeal,fill=nativeTeal!70] coordinates {(GEMM,1.000) (2mm,1.001) (FFN,1.000)};
\addplot[draw=fabricAmber,fill=fabricAmber!35,postaction={pattern=north east lines,pattern color=fabricAmber}] coordinates {(GEMM,0.667) (2mm,0.667) (FFN,0.500)};
\end{groupplot}
\node[anchor=north] at ($(group c2r1.south)+(0,-4.5mm)$) {\ref{streamleg}};
\end{tikzpicture}\par
\vspace{-4mm}
\caption{Stream-HLS benchmarks (OOC placement, 300\,MHz).}
\label{fig:stream}
\end{figure}
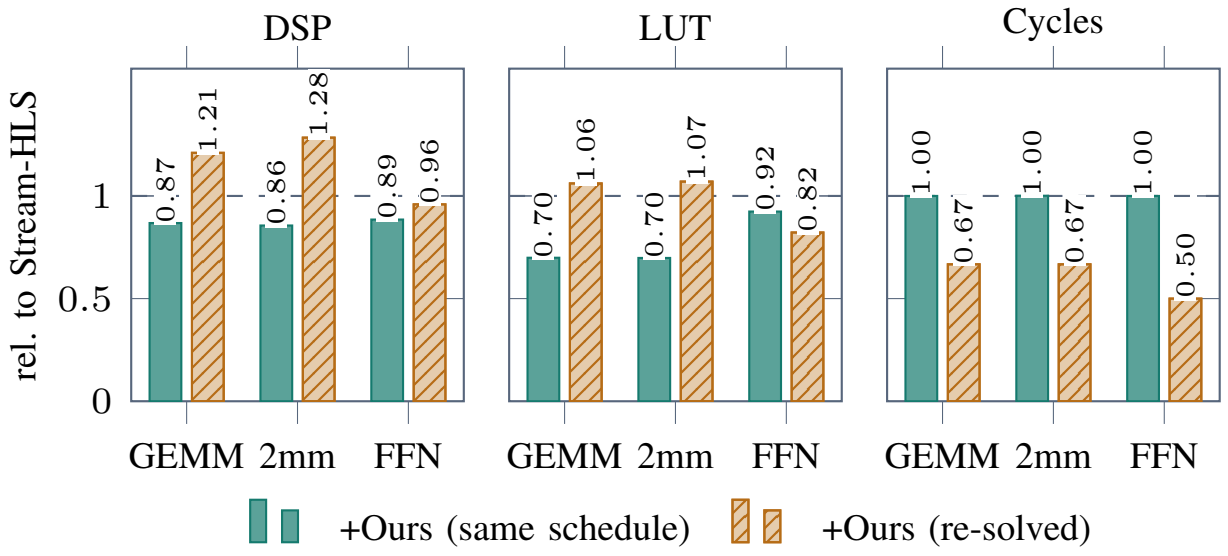

In Stream-HLS, the templates save DSPs under the original schedule, and once the cost model reflects native mapping, the scheduler chooses a faster design. We use three Stream-HLS benchmarks in Fig.~\ref{fig:stream}, namely GEMM, two chained matrix multiplications (2mm), and a transformer feed-forward network block (FFN). We report OOC results, exactly as the original Stream-HLS study does~\cite{streamhls}, so that our baseline is directly comparable with its published setting. First, we keep its task graph and schedule and replace only the arithmetic at the leaves. Stream-HLS lets HLS build adder trees that cost about 1.4 DSPs per MAC, whereas our library builds chains with one DSP per product and one adder per chain. This saves 11--15\% of the DSPs at almost the same cycle count. Then, we re-solve the scheduling model of Stream-HLS with the actual costs of our templates, in which a multiplier feeding an adder counts as one fused multiply--add. The solver now unrolls wider tiles and cuts the cycles by 1.5\X{} in GEMM and 2mm and by 2\X{} in the FFN block. To do so, the solver invests more of the device into wider tiles, more BRAMs and, in GEMM and 2mm, more DSPs and LUTs, while staying within its DSP budget. In the FFN block, it even uses fewer DSPs and LUTs than the original design.

\section{Related Work}
\label{sec:related}

HLS compilers translate algorithmic descriptions into hardware through scheduling, pipelining, and resource binding~\cite{cong2011hls,ug1399}. Higher-level frameworks extend this process. ScaleHLS provides multi-level compiler optimizations~\cite{scalehls2022}, Allo separates computation from composable hardware customizations~\cite{allo2024}, AutoSA generates systolic-array accelerators~\cite{autosa2021}, and Stream-HLS optimizes dataflow schedules and resource allocation~\cite{streamhls}. These frameworks focus on organizing and scheduling computation, whereas our library improves how operators and storage structures map to FPGA hard blocks.

Previous work has explored DSP capabilities beyond conventional arithmetic inference. Arithmetic expression graphs have been mapped onto DSP48E1 blocks, exploiting their internal operations and cascade connections~\cite{ronak2014fpl,ronak2016tcad}. Xilinx WP486 describes packing two INT8 products with a shared operand into a DSP48E2~\cite{wp486}, and DSP-Packing extends arithmetic packing to other operand widths~\cite{dsppacking2022}. Our library extends native mapping across arithmetic modes, internal registers, dedicated interconnections, and memories.

Application-specific frameworks also exploit FPGA hard blocks. FINN and FINN-R generate accelerators for quantized neural networks~\cite{finn2017,finnr2018}, and FINN provides optimized RTL matrix--vector multiplication alongside an HLS implementation in finn-hlslib~\cite{finnrepo,finnhlslib}. We use FINN RTL as the reference for native DSP use and reach its mapping from HLS. Similarly, hls4ml translates neural networks into HLS~\cite{hls4ml2018,aarrestad2021}, and our integration replaces its dot-product computation while preserving the surrounding layer architecture. These comparisons show how native mapping complements specialized accelerator libraries and HLS frameworks.

\section{Conclusion}
\label{sec:conclusion}

We asked whether HLS code can reach the native mapping of FPGA hard blocks without changing the compiler. Our library shows it can for the wide range of configurations we evaluated. C++ templates fix the source patterns, an automatic flow sets the scheduling parameters, and verified RTL patches cover what HLS cannot express. On AMD Versal, our matrix multiplication uses 2.4--4.5\X{} fewer LUTs than AutoSA at equal or higher estimated throughput, and its INT8 matrix--vector units match the three MACs per DSP of FINN RTL. Its templates also improve hls4ml, Allo, and Stream-HLS designs in place. The main lesson is that careful source patterns, scheduling settings, and targeted patches extract native mappings and high performance from HLS, which is conventionally assumed to require hand-written RTL. Future work will support more templates, such as moving INT8 accumulation into the DSPs and extend to other FPGA architectures.

\section*{Content Generated by AI}
ChatGPT and Claude assisted with parts of the code, tables, and figures, and in improving the writing clarity.
\clearpage
\bibliographystyle{IEEEtran}
\bibliography{refs}
\end{document}